\documentclass[a4paper,12pt]{article}

\usepackage{graphics}
\usepackage{graphicx}
\usepackage{epstopdf}
\usepackage{caption}

\usepackage{amssymb}
\usepackage{amsmath}
\usepackage{lscape}

\begin{document}

\title{
\bf On analytical solutions of $f(R)$ modified gravity theories in FLRW cosmologies}

\author{
Silvije Domazet$^{a,}$\thanks{sdomazet@irb.hr}, Voja Radovanovi\'{c}$^{b,}$\thanks{rvoja@ipb.ac.rs},  
Marko Simonovi\'{c}$^{b, c,}$\thanks{marko.simonovic@sissa.it} \\
and Hrvoje \v{S}tefan\v{c}i\'{c}$^{a,}$\thanks{shrvoje@thphys.irb.hr}
}


\vspace{3 cm}
\date{
\centering
$^{a}$Theoretical Physics Division, Rudjer Bo\v{s}kovi\'{c} Institute, \\
   P.O.Box 180, HR-10002 Zagreb, Croatia \\
$^{b}$University of Belgrade, Faculty of Physics \\
Studentski trg 12, 11000 Beograd, Serbia \\
$^{c}$SISSA, via Bonomea 265, 34136 Trieste TS, Italy 
}


\maketitle

\abstract

A novel analytical method for $f(R)$ modified theories without matter in Fried\-mann-Lemaitre-Robertson-Walker spacetimes is introduced. 
The equation of motion for the scale factor in terms of cosmic time is reduced to the equation for the evolution of the Ricci scalar $R$ with the 
Hubble parameter $H$. The solution of equation of motion for actions of the form of power law in Ricci scalar $R$, is presented with a detailed 
elaboration of the action quadratic in $R$. The reverse use of the introduced method is exemplified in finding functional forms $f(R)$ which lead 
to specified scale factor functions. The analytical solutions are corroborated by numerical calculations with excellent agreement. Possible further 
applications to the phases of inflationary expansion and late-time acceleration as well as $f(R)$ theories with radiation are outlined.


\section{Introduction}

The use of General Relativity \cite{gr} enables us to describe and explain many features and observational facts about the Universe we live in 
through enormous periods of time. Standard model of big-bang cosmology is a very good description of the evolution of our Universe. There are of course 
unresolved puzzles, for instance flatness or horizon problems, which could be explained by some version of inflationary model \cite{a2,a4,a7,a8} 
which would provide our Universe with an early-time acceleration. This phase should occur  before the radiation dominated epoch 
of cosmic evolution. Speaking of accelerated phases in the life of the Universe, there is also another one, that is a late-time acceleration 
following a matter dominated epoch \cite{a11,a12,a13,a14,a16}. There are several observational indications of this phase of which the supernova Ia 
results \cite{a17,a18,a19} are probably the most well known example. 
The source of this late-time acceleration is frequently identified with an exotic type of matter/energy called dark energy and there have been many 
ideas and attempts in the literature trying to explain its origin \cite{ Huterer:2000mj,
Copeland:2006wr}. 

In order to accommodate the fact of the late-time acceleration of the cosmic expansion, we need to make some modifications in the standard 
cosmological model. Apart from the already mentioned possibility of the existence of dark energy, a prominent possibility is that 
the gravitational interaction is modified at (at least) cosmic scales.  
One possible way to modify General Relativity are the so called $f(R)$ theories 
\cite{a43,a44,a45,a46,Ottewill,Schmidt89,Muller90,Capozziello93,Schmidt,modgrav,DeFelice:2010aj,Sotiriou:2008rp,Appleby,OdintsovRev,DeLaurentis,CliftonRev}, 
where some function of $R$ is used as a Lagrangian density. Starobinsky \cite{a2} used one of these, 
with a specific form given by  $f(R)=R+ \alpha R^2 $ to propose one of the first inflationary models. 
$f(R)$ theories could possibly describe the late-time acceleration phase, thus providing an effective model for dark energy. 
It is a demanding task to produce such a model without spoiling the successes of the standard big-bang cosmology while at the same time 
satisfying solar system and other astrophysical constraints. 
A number of such models has been proposed with various observable consequences and implications \cite{Turner1,Turner2,Odintsov1,Odintsov2,Odintsov3,Dombriz1,Dombriz2}. 
Modified gravity theories have been also found useful in the resolution of the cosmological constant problem \cite{rel1,rel2,rel3,rel4,rel5,rel6,rel7}.

The principal goal of this paper is to present a novel method for analytical solutions of equations of motion arising in $f(R)$ theories in
FLRW cosmologies without matter. In theories of the $f(R)$ type, the equation of motion is a third order differential equation for the scale factor 
$a$ in terms of cosmic time $t$. The main virtue of the proposed method is that the said third order equation is decomposed into three first order 
ones which can be solved consecutively, i.e. one by one and not simultaneously as a system of first order equations.

The organization of the paper is the following. The first section is the introduction. In the second section the formalism of $f(R)$ theories is briefly 
summarized and analytical approaches and solutions of $f(R)$ theories in the existing literature are discussed. The third section contains the 
description of the novel method. In the fourth section the introduced method is applied to theories $f(R) \sim R^{\alpha}$ with a special emphasis 
on actions quadratic in $R$. The fifth section describes the reverse application of the method: finding the functional form of $f(R)$ which results in a specified 
scale factor function, $a(t)$.
In the sixth section we present a numerical algorithm used for the verification of analytical solutions and in 
the seventh section we present the comparison of analytical and numerical solutions. In the eighth section we describe how to extend our method to $f(R)$ theories 
with radiation.
The paper closes with the conclusions section.    
 

\section{The $f(R)$ modified gravity theories in FLRW cosmologies}

In $f(R)$ theories the gravitational part of the action is 
\begin{equation}
 \label{eq:Sgrav}
S_{\mathrm{grav}}= \frac{1}{16 \pi G} \int d^4 x \sqrt{-g} f(R) \, .
\end{equation}
The variation of the entire action $S=S_{\mathrm{grav}}+S_{\mathrm{matter}}$, where $S_{\mathrm{matter}}$ represents the matter part 
of the action, with respect to the metric $g^{\mu \nu}$ yields equations of motion
\begin{equation}
 \label{eq:eom}
f'(R) R_{\mu \nu} - \frac{1}{2} f(R) g_{\mu \nu} + (\nabla_{\mu} \nabla_{\nu} - g_{\mu \nu} \Box) f'(R)=-8 \pi G T_{\mu \nu} \, .
\end{equation}
In the FLRW metric
\begin{equation}
 \label{eq:metric}
d s^2= d t^2-a(t)^2\left( \frac{d r^2}{1-k r^2} + r ^2 d\Omega^2\right) \, ,
\end{equation}
the $00$ component of (\ref{eq:eom}), for the spatially flat 3D space, then reads
\begin{equation}
 \label{eq:eomFLRW}
3 f'(R) H^2 - \frac{1}{2} (R f'(R)-f(R)) + 3 H \dot{R} f''(R) = 8 \pi G \rho \, .
\end{equation}

Despite widespread numerical methods and computational power nowadays available for finding numerical solutions of the equation of motion (\ref{eq:eomFLRW}), 
the most complete insight into the functioning of any theory is achieved when analytical solutions can be obtained, $f(R)$ theories not being exception to this rule.
There have been many attempts of finding analytical solutions for $f(R)$ theories 
\cite{a43,a44,a45,a46,Ottewill,Schmidt89,Muller90,Capozziello93,Schmidt,modgrav,DeFelice:2010aj,Sotiriou:2008rp,Appleby,OdintsovRev,DeLaurentis,CliftonRev}. In particular, the cosmological power law solutions for the 
scale factor dynamics in universes with and without matter were found in \cite{Cap1,Cap2} (see also \cite{CQG2006BarrowClifton}) for $f(R) \sim R^n$. The existence of G\"{o}del, Einstein and de Sitter 
universes in modified gravity theories was studied in \cite{Barr2} whereas analytical solutions for spherically symmetric systems were found in 
\cite{Barr1,Cliftonspherical}. Methods of finding analytical solutions of $f(R)$ theories using Noether symmetries were presented in \cite{Basilakos}.  

A systematic approach to solving $f(R)$ equations of motion was adopted in \cite{Clifton}. Using a specific scalar field reformulation of $f(R)$ theories, 
Clifton succeeded in finding analytical solutions for $R^{1+\delta}$ theories in vacuum spatially flat FLRW spaces. Furthermore, analytical solutions for 
spatially curved vacuum FLRW spaces and some specific solutions with perfect fluids were also found in \cite{Clifton}. The method used in \cite{Clifton}
requires a number of quite ingenious, but intricate substitutions and changes of variable.

In this paper, we present a general approach to finding analytical solutions in $f(R)$ theories. The proposed method reduces the third order equation 
into three first order ones which can be solved one after the another, i.e. do not have to be solved simultaneously as a system of 
first order equations. We also present solutions of 
$R^{\alpha}$ theories using our approach as a demonstration  of the method. 
The selected models of the type $\sim R^{\alpha}$ were primarily chosen for illustrative purposes, especially because they have so far attracted a lot of 
attention of researchers in the field of modified gravity. These models have been so far mostly ruled out by the observations.
In the light of results of \cite{Clifton}, our results for $R^{\alpha}$ 
theories cannot be considered novel, but their alternative derivation in this paper, simpler and more straightforward in our opinion that that in \cite{Clifton}, 
is new. The greatest virtue of the introduced method is its simplicity and generality.            

In the remainder of the paper we restrict our considerations to spatially flat spaces ($k=0$) without matter, i.e. to vacuum solutions in flat 3D spaces.
Only in section 8 we discuss the addition of radiation in $f(R)$ theories.

\section{The method}

The dynamical equation describing the expansion of the spatially flat universe without matter (or when the contribution of matter is negligible) is
\begin{equation}
 \label{eq:dynfR}
3 F(R) H^2=\frac{1}{2} (F(R) R-f(R)) - 3 H \dot{F}(R) \, ,
\end{equation}
where $F(R)=f'(R)$, prime denotes the differentiation with respect to $R$ and dot represents the differentiation with respect to cosmic time. 
The crucial step of the method presented in this paper stems from the structure of the Ricci scalar in the RW metric:
\begin{equation}
 \label{eq:Ricci}
R=6 \dot{H} + 12 H^2 \, .
\end{equation}
Namely, in intervals of cosmic time in which $H(t)$ is a monotonic function of $t$, it is possible to make the change 
of variables from $t$ to $H$. In particular, 
\begin{equation}
 \label{eq:change}
\frac{d}{dt} = \dot{H} \frac{d}{dH} = \left( \frac{R}{6} - 2 H^2 \right) \frac{d}{dH} \, .
\end{equation}
If $H(t)$ is not a monotonic function of cosmic time $t$, the time domain should be divided into intervals where 
$H(t)$ is a monotonic function and the transformation (\ref{eq:change}) should be performed in each interval separately. 
Inserting (\ref{eq:change}) into (\ref{eq:dynfR}) yields
\begin{equation}
 \label{eq:RodH}
3 H \left( \frac{R}{6} - 2 H^2 \right) f''(R) \frac{dR}{dH} = \frac{1}{2} (f'(R) R-f(R)) - 3 H^2 f'(R) \, .
\end{equation}
The representation of the starting third-order differential equation (\ref{eq:dynfR}) in the form given above opens a way for a program of 
obtaining analytical solutions by solving three first-order differential equations one after the other. In the first step, the solution of (\ref{eq:RodH}) results in the 
function $R(H)$. Once the Ricci scalar $R$ is available as the function of the Hubble expansion rate $H$, solving (\ref{eq:Ricci}) 
results in $H(t)$. Finally, knowing $H(t)$, by the definition of the Hubble expansion rate, leads immediately to $a(t)$. 

Which of the three steps described above can be carried out analytically depends on the very form of the function $f(R)$. 
It should be stressed that the crucial step in (\ref{eq:change}) is not possible for geometries with $k \not = 0$.
The remainder of this paper is devoted to the study of cases where at least one of the steps can be done analytically. 

\section{The applications}

\subsection{Action quadratic in $R$}

\label{quad}

The action quadratic in Ricci scalar is of physical interest for several reasons. 
From a purely formal point of view, it is the power-law term of the lowest positive integer power which yields a nontrivial $f(R)$ modified gravity theory. 
The terms quadratic in $R$ appear as a part of vacuum action in quantum field theory in curved space-time, see e.g. \cite{Shapiro}. 
Finally, $R^2$ terms play a crucial role in one of the most prominent models of inflation, the Starobinsky model \cite{a2} ($R^2$ term is actually a 
truncation of the Starobinsky model $f(R)=R+A R^2$ and it becomes important in the early universe when the curvature is large) and in some recent analyses 
of the radiation epoch of the cosmic evolution \cite{Dolgov}. 

Inserting the function $f(R)=A R^2$, where $A$ is a real constant, into (\ref{eq:RodH}) gives the equation
\begin{equation}
 \label{eq:R2}
H (R-12 H^2) \frac{dR}{dH} = \frac{1}{2} R (R-12 H^2) \, ,
\end{equation}
or more compactly
\begin{equation}
 \label{eq:R2compact}
(R-12 H^2) \left( H \frac{dR}{dH} - \frac{1}{2} R \right)=0 \, .
\end{equation}
As long as $H$ is evolving in time (and $R-12 H^2 \neq 0$) the equation to be solved is 
\begin{equation}
 \label{eq:R2red}
H \frac{dR}{dH}=\frac{1}{2} R \, .
\end{equation}
The solution for $R(H)$ is then
\begin{equation}
 \label{eq:R2sol}
R(H)=R_0 \left( \frac{H}{H_0} \right)^{1/2} \, ,
\end{equation}
where $R(H_0)=R_0$ is the initial condition.

Eq. (\ref{eq:Ricci}) now acquires the form
\begin{equation}
 \label{eq:R2Hdot}
\frac{dH}{dt}=\frac{R_0}{6} \left( \frac{H}{H_0} \right)^{1/2} - 2 H^2
\end{equation}
with the following closed-form solution:
\begin{eqnarray}
 \label{eq:R2Hodtsol}
H_0 (t-t_0) &=& -\frac{1}{6 \chi^{2/3}} \ln \frac{(\chi^{1/3} - (H/H_0)^{1/2})^2}{\chi^{2/3}+\chi^{1/3} (H/H_0)^{1/2} + (H/H_0)} \nonumber \\ &+& \frac{1}{\chi^{2/3}\sqrt{3}} \arctan \frac{2 (H/H_0)^{1/2} + \chi^{1/3}}{\chi^{1/3} \sqrt{3}} \nonumber \\
&+& \frac{1}{6 \chi^{2/3}} \ln \frac{(\chi^{1/3} - 1)^2}{\chi^{2/3}+\chi^{1/3} +1} -  \frac{1}{\chi^{2/3}\sqrt{3}} \arctan \frac{2 + \chi^{1/3}}{\chi^{1/3} \sqrt{3}} \, ,
\end{eqnarray}
where $\chi=R_0/(12 H_0^2)$. This solution implicitly defines $H$ in terms of $t$. From the definition of the Hubble parameter, $H=\dot{a}/a$, one 
readily obtains the equation for the evolution of $a$ with $H$. Namely, from (\ref{eq:R2Hdot}) it follows
\begin{equation}
 \label{eq:aodH}
\frac{d a}{a} = \frac{H \, dH}{\dot{H}}=\frac{ H \, dH}{\frac{R_0}{6} \left(\frac{H}{H_0} \right)^{1/2}-2 H^2} \, .
\end{equation}
A straightforward integration of this equation yields
\begin{equation}
 \label{eq:asol}
a=a_0 \left[ \frac{\chi - (H/H_0)^{3/2}}{\chi-1} \right]^{-1/3} \, ,
\end{equation}
whereas an inversion of this expression yields 
\begin{equation}
\label{eq:Hoda}
H=H_0 \left[\chi - (\chi-1) \left( \frac{a}{a_0} \right)^{-3} \right]^{2/3} \, .
\end{equation}
The expression (\ref{eq:Hoda}) was already obtained in \cite{BarrowQuadratic} using a different method. 
This solution can be inserted into (\ref{eq:R2Hodtsol}) to obtain the implicit relation connecting $t$ and $a$. Finally, 
from the solutions (\ref{eq:R2Hodtsol}) and (\ref{eq:Hoda}) one can easily read out the asymptotic behavior: for $t \rightarrow \infty$, the scale factor is 
unbounded, $a \rightarrow \infty$, whereas the Hubble parameter saturates, $H \rightarrow \chi^{2/3} H_0$.

\subsection{Separability of $R-12 H^2$ term}

The effectiveness of the method for the action quadratic in $R$ is closely connected to a very simple form of equation for $R$ as a function of $H$, Eq. (\ref{eq:R2red}). This simple form is a consequence of the factorization of the $R-12 H^2$ term or the right-hand side of (\ref{eq:RodH}), as explicitly demonstrated in (\ref{eq:R2compact}). An interesting question is which other, if any, functional forms $f(R)$ also allow the factorization of the $R-12 H^2$ term on the right-hand side of (\ref{eq:RodH}). Indeed, the factorization 
\begin{equation}
 \label{eq:factor}
\frac{1}{2} (f'(R) R-f(R)) - 3 H^2 f'(R)=\tau(H) \lambda(R) (R-12 H^2) \, ,
\end{equation}
where $\tau(H)$ and $\lambda(H)$ are arbitrary functions of $H$ and $R$, respectively, results in a separable equation once the $R-12 H^2$ term is factored out. Term by term comparison of right and left-hand side in (\ref{eq:factor}) gives $\tau(H)=const$ which can be absorbed into $\lambda(R)$ so that we can take $\tau(H)=1$. Further we obtain
\begin{equation}
 \label{eq:comp1}
\lambda(R)=\frac{1}{4} f'(R) 
\end{equation}
and 
\begin{equation}
 \label{eq:comp2}
\frac{1}{2} (f'(R) R-f(R))=\lambda(R) R \, .
\end{equation}
Combination of (\ref{eq:comp1}) and (\ref{eq:comp2}) gives $f(R)=A R^2$. Therefore, the action quadratic in $R$ is the only choice for $f(R)$ which allows factorization as described in (\ref{eq:factor}). 

\subsection{$f(R)=A R^{\alpha}$}

The general power law form $f(R)=A R^{\alpha}$ is a natural extension of the quadratic action. 
This choice of action has recently raised a lot of attention in the literature as adding such a power law term 
to the $R$ term in the modified gravity action 
can lead to accelerated late-time expansion of the universe such as for negative values of $\alpha$.
The action $f(R)=A R^{\alpha}$ is chosen primarily for illustration purposes. It should be stressed that it is strongly constrained \cite{CQG2006BarrowClifton,Barr1} 
and disfavored as a model of late-time universe by light-deflection and planetary period observations \cite{Zakharov} and the requirement of avoiding the Dolgov-Kawasaki
instability \cite{DolgovKawasaki}.

For this choice of $f(R)$ Eq. (\ref{eq:RodH}) acquires the form
\begin{equation}
 \label{eq:power}
\alpha (\alpha-1) (R-12 H^2) H \frac{dR}{dH} = (\alpha-1)R^2-6 \alpha H^2 R \, .
\end{equation}
Using the substitution $R=\xi H^2$, the equation above can be transformed into a separable form
\begin{equation}
 \label{eq:xi}
\frac{dH}{H}=\frac{\alpha}{1-2\alpha} \frac{\xi-12}{\xi (\xi+\beta)} d\xi \, ,
\end{equation}
with 
\begin{equation}
 \label{eq:beta}
\beta = \frac{6 \alpha (4 \alpha -5)}{(\alpha-1)(1-2\alpha)} \, .
\end{equation}

Writing

\begin{equation}
 \frac{\xi-12}{\xi (\xi+\beta)}=\frac{C_1}{\xi}+\frac{C_2}{\xi+\beta}=\frac{(C_1+C_2)\xi+C_1\beta}{\xi(\xi+\beta)} \, ,
\end{equation}


we can see that

\begin{equation}\begin{split}
&C_1+C_2=1 \, ,\\
&C_1=-\frac{12}{\beta} \, ,\\
&C_2=1+\frac{12}{\beta} \, .\\
\end{split}\end{equation}



Equation (\ref{eq:xi}) can be readily integrated to obtain the solution in closed form: 
\begin{equation}
 \label{eq:powersol}
\left( \frac{H}{H_0} \right)^{(1-2\alpha)/\alpha} = \left( \frac{\xi}{\xi_0} \right)^{-12/\beta} \left( \frac{\xi+\beta}{\xi_0+\beta} \right)^{1+12/\beta} \, .
\end{equation}

In general, Eq. (\ref{eq:powersol}) can be considered as a parametric solution of the $R(H)$ relation. Namely, (\ref{eq:powersol}) yields $H(\xi)$, 
whereas $R(\xi)$ is obtained directly from its definition as $R(\xi)=\xi H(\xi)^2$.

Let us further consider some values of $\alpha$ for which (\ref{eq:powersol}) can be inverted to obtain the explicit relation $R=R(H)$. Defining
\begin{equation}
K=\frac{\xi_0^{C_1}(\xi_0+\beta)^{C_2}}{H_0^{\frac{1-2\alpha}{\alpha}}} \, ,
\end{equation}
(\ref{eq:powersol}) can be written as
\begin{equation}
\label{K}
\xi^{C_1}(\xi+\beta)^{C_2}=KH^{\frac{1-2\alpha}{\alpha}} \, .
\end{equation}
Let us examine the following combinations of $C_1$ and $C_2$ (and therefore values of $\alpha$):

\begin{enumerate}
\item $C_1=-2C_2$  will give $C_2=-1$ and $C_1=2$ and the equation (\ref{K}) becomes

\begin{equation}
\xi^{2}(\xi+\beta)^{-1}=KH^{\frac{1-2\alpha}{\alpha}} \, ,
\end{equation}

with solutions
\begin{equation}
\xi_{1,2}=\frac{1}{2}KH^{\frac{1-2\alpha}{\alpha}}\left(1\pm\sqrt{1-24K^{-1}H^{-\frac{1-2\alpha}{\alpha}}}\right) \, .
\end{equation}

This leads to solutions for $R$

\begin{equation}
R_{1,2}=\frac{K}{2}H^{\frac{1}{\alpha}}\left(1\pm\sqrt{1-24K^{-1}H^{-\frac{1-2\alpha}{\alpha}}}\right) \, ,
\end{equation}

where 

\begin{equation}
\alpha_{1,2}=\frac{1\pm\sqrt{3}}{2} \, .
\end{equation}

\item $C_2=-2C_1$  will give $C_1=-1$ and $C_2=2$ so this time (\ref{K}) becomes

\begin{equation}
\xi^{-1}(\xi+\beta)^{2}=KH^{\frac{1-2\alpha}{\alpha}} \, ,
\end{equation}

leading to solutions

\begin{equation}
\xi_{1,2}=\frac{1}{2}\left(-24+KH^{\frac{1-2\alpha}{\alpha}}\pm\sqrt{\left(KH^{\frac{1-2\alpha}{\alpha}}-48\right)KH^{\frac{1-2\alpha}{\alpha}}}  \right)
\, ,
\end{equation}


so that

\begin{eqnarray}
R_{1,2}=\frac{1}{2}H^{2}\left(-24+KH^{\frac{1-2\alpha}{\alpha}}\pm\sqrt{\left(KH^{\frac{1-2\alpha}{\alpha}}-48\right)KH^{\frac{1-2\alpha}{\alpha}}}\right)
\, , 
\end{eqnarray}
where
\begin{equation}
\alpha_{1,2}=\frac{11\pm\sqrt{57}}{16} \, .
\end{equation}

\item Another case which would lead to quadratic equation in $\xi$ would be  $C_1=C_2=\frac{1}{2}=-\frac{12}{\beta}$ where we would have 
$\beta=-24$ which would lead to a complex values of $\alpha$:
\begin{equation}
\alpha_{1,2}=\frac{7\pm\sqrt{-15}}{8} \, .
\end{equation}

\end{enumerate}

Finally, the inspection of (\ref{eq:powersol}) and the definitions of $C_1$, $C_2$ and $\beta$, reveal the following characteristic intervals and point values 
of $\alpha$: $(\infty, 0)$, $[0,1/2)$, $1/2$, $(1/2,1)$, $[1,5/4)$, $5/4$, $(5/4,2)$ and $[2,\infty)$. 
 For points of $\alpha=1/2,5/4$ we solve (\ref{eq:power}) directly by inserting concrete values of $\alpha$.
This results in

\begin{equation}
H=H_0 e^{\frac{1}{36}(\xi-\xi _0)}\left(\frac{\xi}{\xi _0}\right)^{-\frac{1}{3}} \, ,
\end{equation}
and 
\begin{equation}
H=H_0 e^{10\frac{(\xi-\xi _0)}{\xi \xi _0}}\left(\frac{\xi}{\xi _0}\right)^{-\frac{5}{6}} \, ,
\end{equation}
for  $\alpha=1/2,5/4$ respectively, with $R=\xi H^2$. 

\section{Functional form of $f(R)$ from the known solutions for the scale factor}

The preceding section was dedicated to finding analytical solutions for the known functional form $f(R)$. The ordering of this approach can also be reversed. 
Namely, for a known scale factor dependence on time, $a(t)$, one may ask which functional forms $f(R)$ yield such $a(t)$ functions as solutions. 
We consider the reconstruction in the mathematical sense, i.e. obtaining the function $f(R)$ which without matter leads to a known function $a(t)$.
The procedure is the following: from a known $a(t)$ calculate $\dot{a}(t)$, then $H(t)$ and $\dot{H}(t)$. Using (\ref{eq:Ricci}), the expression for $R(t)$ follows 
immediately. If from expressions for $H(t)$ and $R(t)$ it is possible to eliminate $t$ and obtain relation $H^2=k(R)$, the relation (\ref{eq:RodH}) becomes a first order 
differential equation for the function $f(R)$. For related work on the reconstruction methods in $f(R)$ gravity see 
\cite{reconstruction1,reconstruction2,reconstruction3,rec1,rec2,recRT}.

As an illustration of the procedure explained above, we determine functional forms $f(R)$ which lead to 
power law expansion, $a(t) \sim t^{\beta}$ and singular future behavior $a(t) \sim 1/(T-t)^m$. 

%

\subsection{Power law expansion}   

For the power law expansion
\begin{equation}
\label{eq:pow1}
a(t)=D t^{\beta} \, , 
\end{equation}
where $D$ and $\beta$ are constants, we have $H(t)=\beta/t$ and $\dot{H}(t)=-\beta/t^2$. The expression for the Ricci scalar is $R(t)=6 \beta (2 \beta-1)/t^2$ which 
is readily displayed as $R=\frac{6 (2 \beta-1)}{\beta} H^2 \equiv \frac{1}{\gamma} H^2$. Eq. (\ref{eq:RodH}) now acquires the form
\begin{equation}
 \label{eq:pow2}
(1-12 \gamma) R^2 f''(R) - \frac{1}{2} (1-6 \gamma) R f'(R) + \frac{1}{2} f(R) = 0 \, .
\end{equation}
This equation has solutions of the form $f(R) \sim R^{\lambda}$ where $\lambda$ is the solution of the equation 
\begin{equation}
 \label{eq:pow3}
2 \lambda^2+(\beta-3) \lambda + 1-2\beta=0 \, ,
\end{equation}
in accordance with the results of \cite{Cap1,Cap2}. Direct inspection of this equation shows that $\lambda=2$ is not its solution for any value of $\beta$. 
Solving (\ref{eq:pow3}) follows
\begin{equation}
 \label{eq:pow4}
\lambda_{1,2}=\frac{1}{4} (3-\beta \pm \sqrt{\beta^2+10 \beta +1}) \, .
\end{equation}
In an expanding universe $\beta>0$ and $\lambda_1 \not = \lambda_2$.
This result shows that the functional form $f(R)$ that leads to the power law expansion (\ref{eq:pow1}) 
is a linear combination of $R^{\lambda_1}$ and $R^{\lambda_2}$. See also \cite{power} for a similar result with matter.

\subsection{Future singularities}

As some mechanisms for the explanation of the late-time accelerated expansion of the universe lead to singularities in finite time 
in the future \cite{OdintsovRev} (phantom energy with the constant equation of state parameter being one such example), study of such expansion scenarios is of interest in modified
gravity theories as well. For a scale factor evolution with a future singularity at time $T > t_0$
\begin{equation} 
 \label{eq:sing1}
a=\frac{A}{(T-t)^m} \, ,
\end{equation}
where $A$ and $m$ are positive constants, the Hubble parameter and its derivative have the form 
\begin{equation}
 \label{eq:sing2}
H=\frac{m}{T-t} \, , \;\;\;\; \dot{H}=\frac{m}{(T-t)^2} \, .
\end{equation}
The Ricci scalar is again quadratically dependent on $H$:
\begin{equation}
 \label{eq:sing3}
R=\frac{6(2 m +1)}{m} H^2 \, .
\end{equation}
Using this result (\ref{eq:RodH}) becomes
\begin{equation}
 \label{eq:sing4}
2R^2f^{\prime\prime}(R)-(m+1)Rf^\prime(R)+(2m+1)f(R)=0 \, .
\end{equation}
The solutions of this equation are in the $R^{\lambda}$ form where the exponents $\lambda$ are
\begin{equation}
 \label{eq:sing5}
\lambda_{1,2}=\frac{m+3\pm\sqrt{m^2-10m+1}}{4} \, .
\end{equation}
For $ m < m_1= 5-2\sqrt{6}$ and $m > m_2=5+2\sqrt{6}$, both solutions for $\lambda$ are real and the solution for $f(R)$ is
\begin{equation}
 \label{eq:sing6}
f(R)=K_1 R^{\lambda_1}+ K_2 R^{\lambda_2} \, ,
\end{equation}
where $K_{1,2}$ are real constants. For $m=m_{1,2}$ the solution for $f(R)$ is a linear combination of
$R^{(m_{1,2}+3)/4}$ and $R^{(m_{1,2}+3)/4} \ln R$. Finally, for $m_1< m < m_2$ the solutions for $\lambda$ are complex conjugate, 
$\lambda_{1,2}=\lambda_R \pm \lambda_I$, with $\lambda_R=\frac{m+3}{4}$ and $\lambda_R=\frac{1}{4}\sqrt{|m^2-10 m+1|}$. The solution for $f(R)$ leading to
(\ref{eq:sing1}) is
\begin{equation}
 \label{eq:sing7}
f(R)=K_3 R^{\lambda_R} \cos (\lambda_I \ln R) + K_4 R^{\lambda_R} \sin (\lambda_I \ln R) \, ,
\end{equation}
where $K_{3,4}$ are real constants. The obtained results are consistent with \cite{OdintsovRev}.


The examples discussed in this section illustrate one more additional advantage of the analytical method introduced in this paper. 
It allows us to find some $f(R)$ actions leading to predefined function $a(t)$. Thus,
it could potentially add to the existing literature on the reconstruction of modified gravity models 
which mimic known expansion epochs such as inflation, matter dominated or radiation dominated epoch of the expansion of the universe.

\section{Numerical solutions}

When we have knowledge of the $f(R)$ function, we want to find the evolution of the scale factor and other 
quantities (Hubble parameter, curvature\ldots). As we have seen in the above sections, unlike standard Friedmann equations, 
analogous equations for arbitrary $f(R)$ are very complicated and hard to solve. When the matter is absent the equation requiring solution is

\begin{equation}
3F(R)\left(\frac{\dot{a}}{a}\right)^2=\frac{1}{2}\left(F(R)R-f(R)\right)-18\frac{\dot{a}}{a}\frac{dF}{dR}\left(\frac{\dddot{a}}{a}-
\frac{\dot{a}\ddot{a}}{a^2}-2\left(\frac{\dot{a}}{a}\right)^3\right) \, .
\end{equation}

It can be rewritten in a more suitable form for numerical integration as a system of three differential equations of the first order

\begin{equation}
\dot{a}=b \, ,
\end{equation}

\begin{equation}
\dot{b}=\ddot{a}=c\, ,
\end{equation}

\begin{equation}
\label{eq:cdot}
\dot{c}=\dddot{a}=\frac{a^2}{18b\frac{dF}{dR}}\bigg(\frac{1}{2}(FR-f)-3F\frac{b^2}{a^2}\bigg)-\frac{b}{a}\left(c-\frac{2b^2}{a}\right) \, .
\end{equation}

For a general class of $f(R)$ functions

\begin{equation}
\label{eq:fr}
f(R)=\beta R+\alpha (R-R_0)^n \, ,
\end{equation}
the parameters $\alpha$, $\beta$, $ R_0$ and $n$ can be chosen in such a way to recover some theories where the solutions are known. 
The only restriction is not to go to standard GR plus cosmological constant limit, because then we would have $n=0$ or $n=1$ or $\alpha=0$ 
because this would lead to divergences in (\ref{eq:cdot}). Parameter $R_0$ serves to avoid possible divergences in the $n<0$ case.

For $f(R)$ chosen as in (\ref{eq:fr}) equation (\ref{eq:cdot}) becomes 

\begin{eqnarray}
\label{eq:cdot2}
\dot{c}=\dddot{a}&=&\frac{a^2}{18b}\frac{1}{an(n-1)}\frac{1}{(R-R_0)^{n-2}}\bigg(\frac{n-1}{2}\alpha (R-R_0)^n- \nonumber \\&-&
3\left( \beta+\alpha n(R-R_0)^{n-1}\right)\frac{b^2}{a^2}\bigg)-\frac{b}{a}\left(c-\frac{2b^2}{a}\right) 
\end{eqnarray}

The output is given in terms of  
$a(t)$, $b(t)$, $c(t)$ functions so that any other quantity we need can be calculated, such as $R(t)$, $H(t)$, or $q(t)$. This allows 
analysis of functional dependencies, such as $R(H)$.

We need to make our variables dimensionless. To accomplish that take $\tau=H_0(t-t_0)$ leading to $d\tau=H_0dt$. 
Also let $a=a_0x$ where $a_0$ is the scale parameter today. Then

\begin{eqnarray}
&\dot{a}=a_0\dot{x}=a_0H_0\frac{dx}{d\tau} \, ,\nonumber \\ 
&\ddot{a}=a_0H_0^2\frac{d^2x}{d\tau^2} \, .
\end{eqnarray} 

Any other quantity is scaled accordingly, so for example

\begin{eqnarray}
&H=H_0 \frac{1}{x} \frac{dx}{d\tau} \, ,\nonumber \\
&R=6H_0^2\left[\left(\frac{1}{x} \frac{dx}{d\tau}\right)^2+\frac{1}{x} \frac{d^2x}{d\tau^2}\right]
\end{eqnarray}
and dimensionless quantities can be constructed

\begin{eqnarray}
h=\frac{H}{H_0}\, , && r=\frac{R}{H_0^2}\, .
\end{eqnarray}

The graphs representing numerical solutions generated by this program along with analytical solutions for several different choices of 
$f(R)=R^{\alpha}$ are discussed in the following section.

\section{Results and discussion}

\begin{figure}[h!]
\centering
\includegraphics[width=8cm]{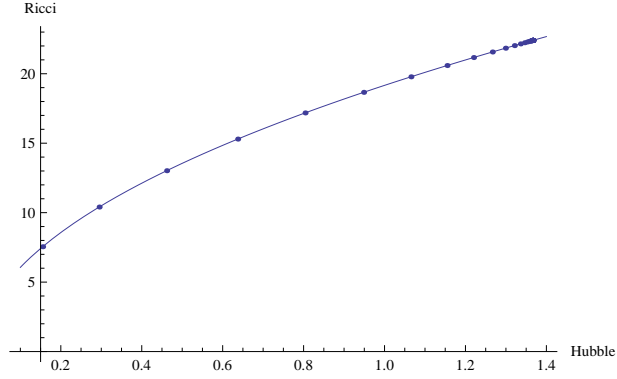} 
\captionsetup{font=scriptsize}
\caption[Rpower2]{\label{fig:f1} Functional dependence of dimensionless quantities $r(h)$ for $f(R)=R^2$} 
\end{figure}

\begin{figure}[h!]
\centering
\includegraphics[width=7cm]{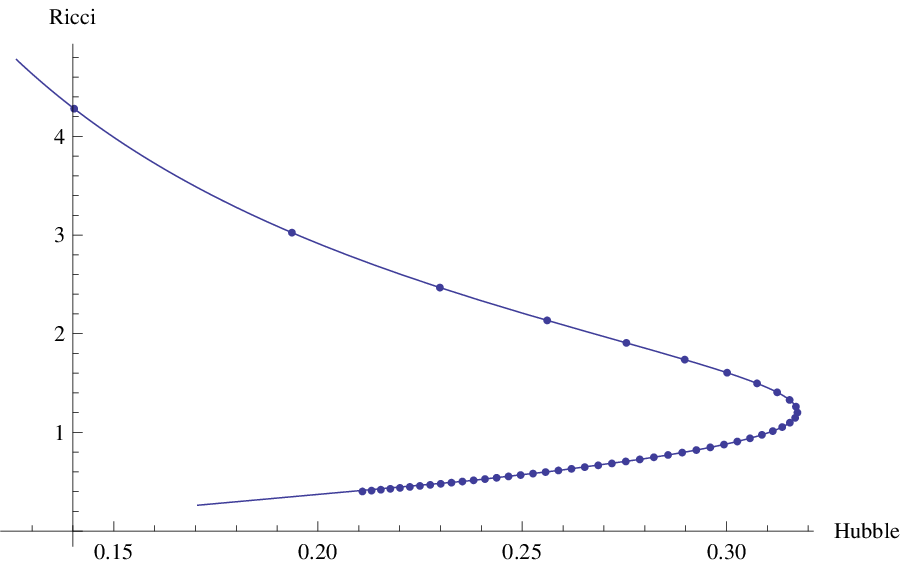}
\includegraphics[width=7cm]{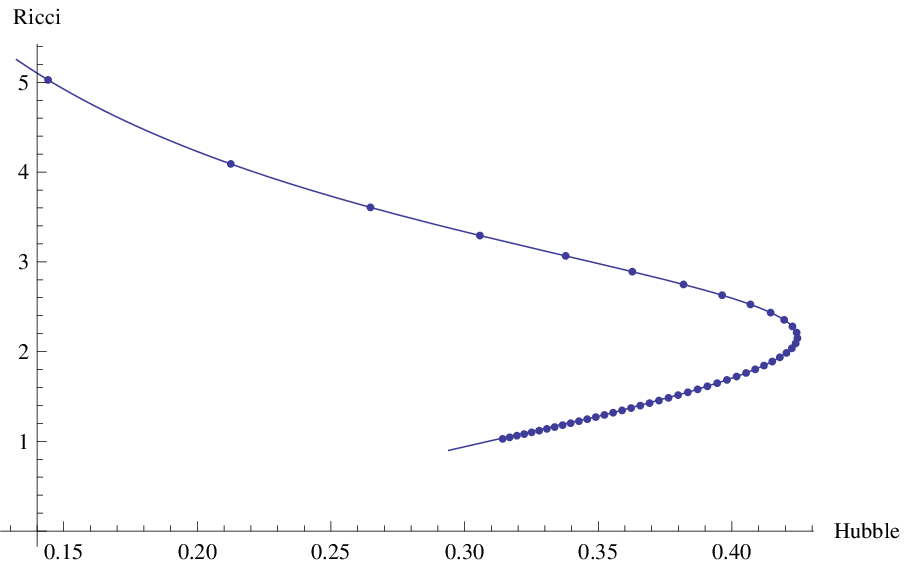}
\captionsetup{font=scriptsize}
\caption[NegativeRpower]{\label{fig:f2} $r(h)$ dependence for   $f(R)=R^{-1}$ and  $f(R)=R^{-2}$}
\end{figure}

\begin{figure}[h!]
\centering
\includegraphics[width=7cm]{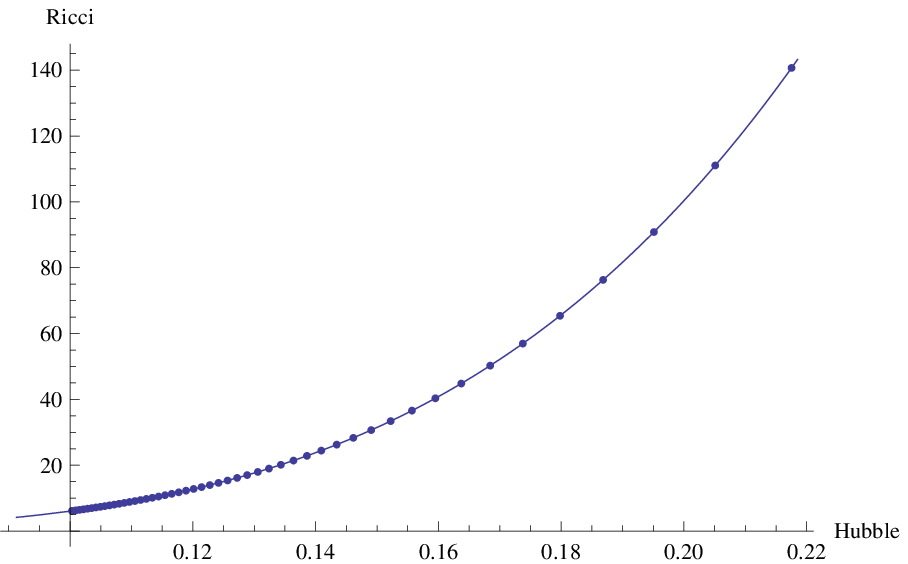}
\includegraphics[width=7cm]{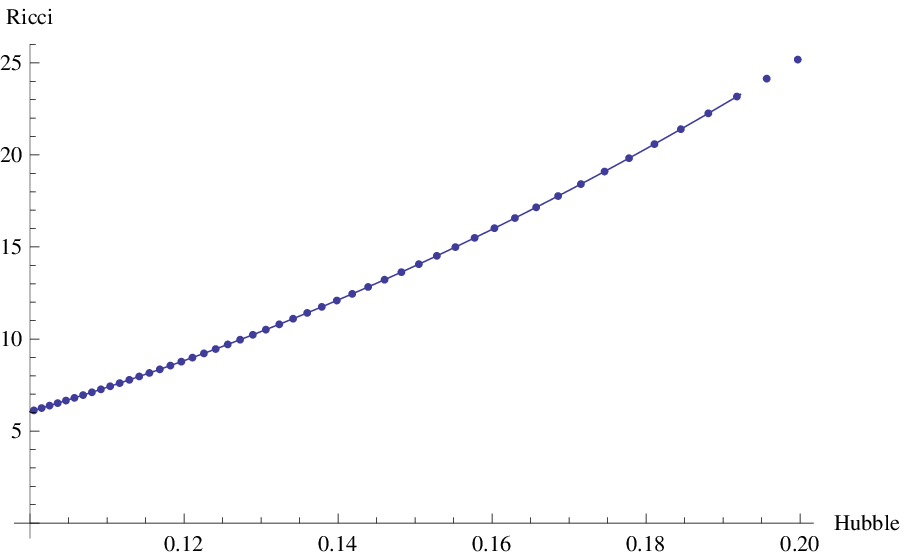}
\captionsetup{font=scriptsize}
\caption[FractionalRpower]{\label{fig:f3} $r(h)$ dependence for   $f(R)=R^{0.25}$ and  $f(R)=R^{0.5}$}
\end{figure}

\begin{figure}[h!]
\centering
\includegraphics[width=7cm]{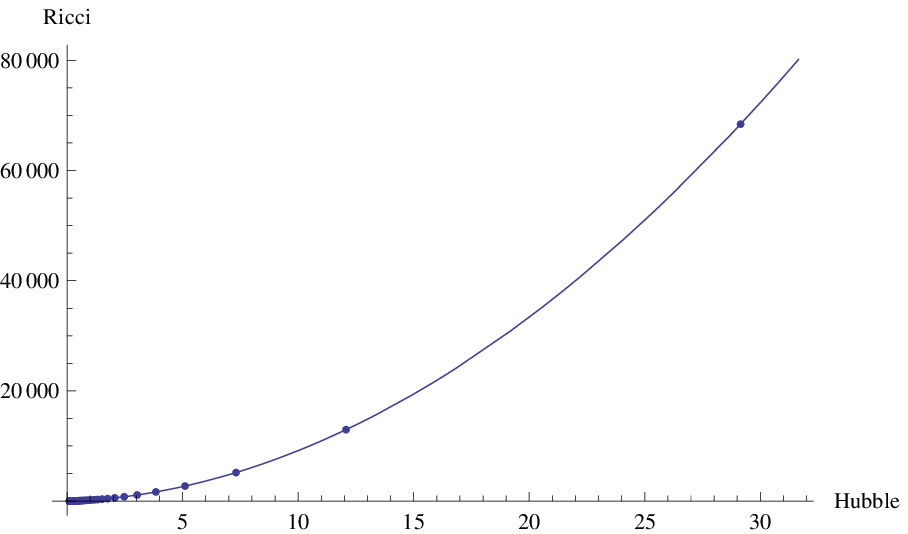}
\includegraphics[width=7cm]{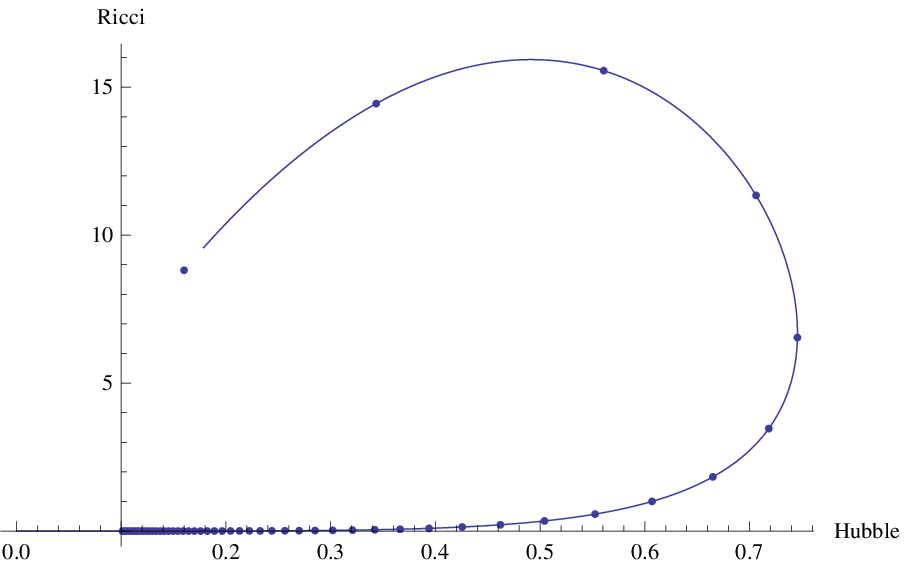}
\captionsetup{font=scriptsize}
\caption[PositiveRpower]{\label{fig:f4} $r(h)$ dependence for   $f(R)=R^{0.75}$ and  $f(R)=R^{1.1}$}
\end{figure}

\begin{figure}[h!]
\centering
\includegraphics[width=7cm]{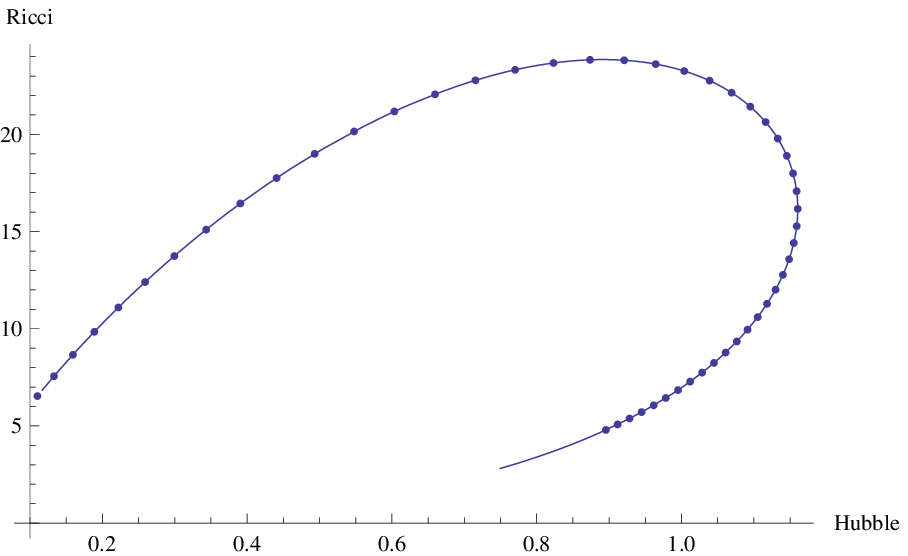}
\includegraphics[width=7cm]{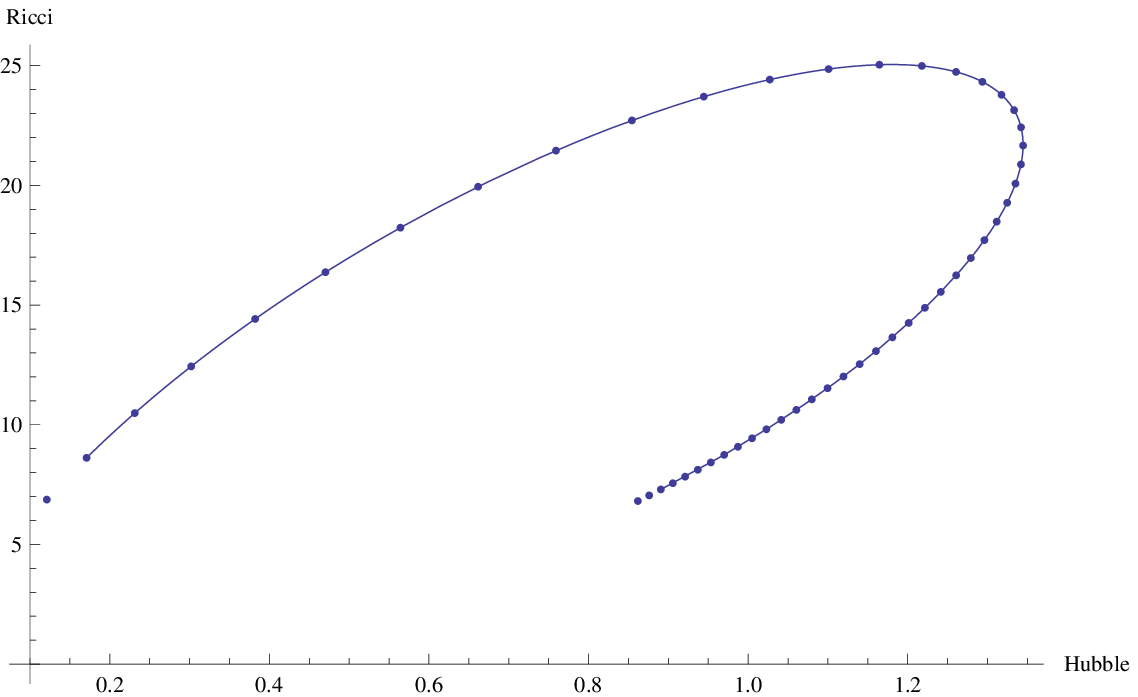}
\captionsetup{font=scriptsize}
\caption[PositiveRpower]{\label{fig:f5} $r(h)$ dependence for   $f(R)=R^{5/4}$ and  $f(R)=R^{1.5}$}
\end{figure}

\begin{figure}[h!]
\centering
\includegraphics[width=7cm]{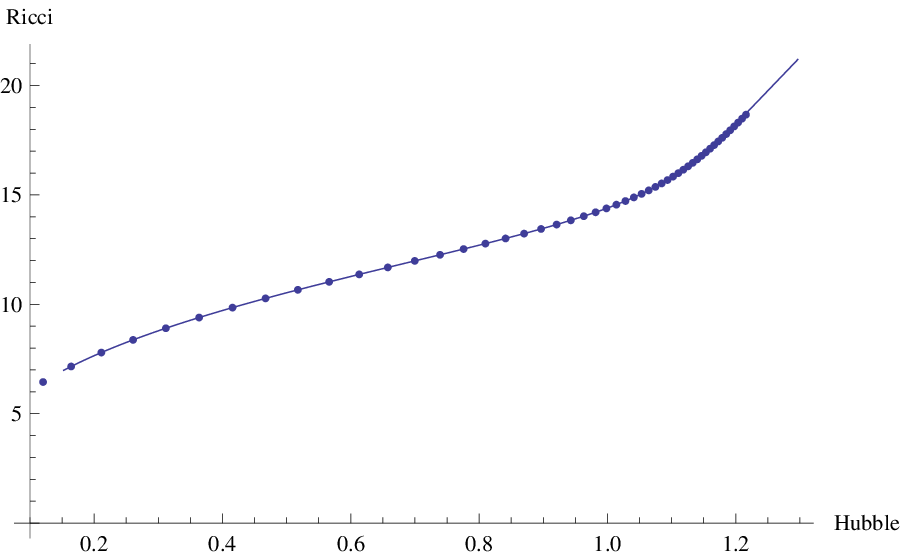}
\includegraphics[width=7cm]{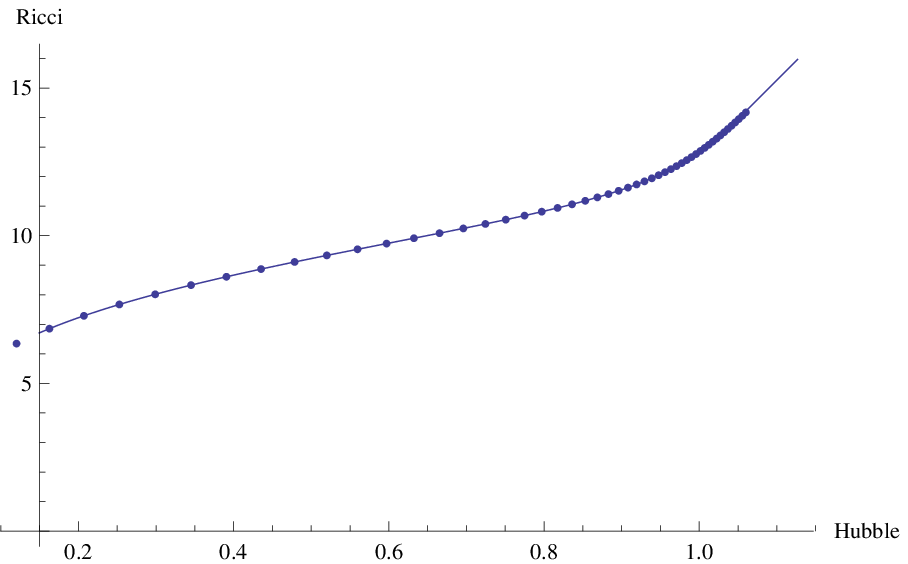}
\captionsetup{font=scriptsize}
\caption[PositiveRpower]{\label{fig:f6} $r(h)$ dependence for   $f(R)=R^3$ and  $f(R)=R^4$}
\end{figure}

In this section we verify our analytical solutions for $R^{\alpha}$ theories obtained in section 4 by comparing them with numerical solutions obtained 
using algorithm laid out in section 6. The results of the comparison for all characteristic point values and intervals for $\alpha$ are presented in 
Figures \ref{fig:f1}-\ref{fig:f6}. 
The line in the figures represents a parametric plot of $R(\xi)$-$H(\xi)$  which is obtained using (\ref{eq:powersol}) and $R=\xi H^2$ 
while the dots represent the output of the numerical procedure.
We reiterate that it is important to apply our analytical method only in intervals of cosmic time in which $H(t)$ is a monotonic function. 

A distinctive feature of some of the plots presented in Figures \ref{fig:f1}-\ref{fig:f6} is the existence of extrema of functions $H(\xi)$ and/or 
$R(\xi)$. In order to analise the extrema of $R(\xi)$-$H(\xi)$ functions we write 

\begin{equation}
\label{eq:hdep}
H=A\xi^{\gamma _1 }(\xi + \beta)^{\gamma _2} 
\end{equation}

and 

\begin{equation}
\label{eq:rdep}
R=H^2 \xi=B\xi^{2\gamma _1+1}(\xi + \beta)^{2\gamma _2} \, ,
\end{equation}
where $\gamma_1=-2 \frac{\alpha-1}{4 \alpha-5}$ and $\gamma_2=\frac{\alpha-2}{(4\alpha-5)(1-2\alpha)}$.
From these two equations we have 

\begin{equation}
\label{eq:rhdep}
\frac{dR}{dH}=\frac{\frac{dR}{d\xi}}{\frac{dH}{d\xi}}=\frac{B}{A}\xi^{\gamma _1 +1}(\xi + \beta)^{\gamma _2}   \frac{(2\gamma _1 +1)(\xi + \beta)+2\xi \gamma _2}{\gamma _1(\xi +\beta)+\xi \gamma _2 }
\, . 
\end{equation}

Looking at this expression we can see that $R$ has an extremum for $\xi=0$ if 
$\gamma _1+1>0$ (equivalently $\alpha \in (-\infty, 5/4) \bigcup (3/2, \infty)$ )  and the same goes for $H$  if  
$\gamma _1+1<0$ (for $\alpha \in (5/4, 3/2)$). 
Also at $\xi =-\beta$ for $\gamma _2>0$ (equivalently for $\alpha \in (-\infty,1/2) \bigcup (5/4,2)$) $R$ has an extremum 
while $H$ has an extremum for $\gamma _2<0$ (for $\alpha \in (1/2, 5/4) \bigcup (2,\infty)$). Also the analysis of possible 
extrema coming from the numerator and denominator of the fraction in Eq. (\ref{eq:rhdep}) one can see that the condition for $R$ to have an 
extremum is $\xi=\frac{6\alpha}{\alpha-1}$ and for the extremum in $H$ condition is $\xi =12$.

\vspace{1cm}

\section{$f(R)$ theories with radiation}

The method of solving $f(R)$ theories can be extended beyond vacuum spaces. To do that, it is necessary to go beyond the $00$ component equation of motion.
The trace of the equation of motion (\ref{eq:eom}) yields a relation
\begin{equation}
 \label{eq:EOMtrace}
R f'(R) - 2 f(R) - 3 \Box f'(R) = - 8 \pi G T \, ,
\end{equation}
where $T \equiv T^{\mu}_{\mu}$ is the trace of the energy-momentum tensor of matter. In the FLRW cosmologies the d'Alambertian operator can be represented as
\begin{equation}
 \label{eq:dAlambertian}
\Box=\frac{d^2}{d \, t^2}+3 H \frac{d}{d \, t} \, .
\end{equation}
In combination with the expression $\frac{d}{dt}=(\frac{R}{6}-2 H^2) \frac{d}{d H}$, it is evident that in FLRW the d'Alambertian operator can be expressed in 
terms of derivatives with respect to $H$. On the other hand, if the matter is dominated by radiation, the trace of energy-momentum tensor $T=\rho_r-3 p_r$ vanishes.
The equation of motion for $f(R)$ theory with radiation can therefore be represented as
\begin{equation}
 \label{eq:EQrad}
R f'(R) - 2 f(R) - 3 \left( \frac{R}{6}-2 H^2 \right) \frac{d}{d H} \left[\left( \frac{R}{6}-2 H^2 \right) \frac{d f'(R)}{d H} \right] - 
9 H \left( \frac{R}{6}-2 H^2 \right) \frac{d f'(R)}{d H} = 0 \, .
\end{equation}
As in the vacuum case, the problem is reduced to three independent differential equations which can be solved consecutively. The case with radiation, however, differs 
from the vacuum case in the fact that the differential equation for $R(H)$ is of second order. 

Apart from a direct application on $f(R)$ theories with radiation, Eq. (\ref{eq:EQrad}) is also applicable in the case of no matter, since $T=0$ in that case too. 
The method represented by Eq. (\ref{eq:EQrad}) may, therefore, also be applied in the study of the transition from the vacuum case to the radiation case or vice versa, 
should this case appear in some $f(R)$ model. The transition from the epoch of inflation to the epoch of radiation might be the first place where the potential of 
(\ref{eq:EQrad}) might be tested.

\section{Conclusions}

We have presented a novel analytical approach to solving vacuum equations of motion for $f(R)$ theories in FLRW spaces. A key advantage of the method 
is the decomposition of the third-order differential equation into three first-order ones which can be solved consecutively. Compared to methods used 
to obtain some exact solutions in the literature \cite{Clifton}, the approach introduced in this paper is simple and much more universal. 
The introduced method also provides an approach to finding functional forms $f(R)$ which produce a given dependence of the scale factor on cosmic time. This fact
paves the way for the use of the method in building $f(R)$ models which closely mimic some important phases in the evolution of the universe and which 
could hopefully be useful in tackling problems of dark matter and dark energy. We also outline the application of the method to $f(R)$ theories with radiation.
As vacuum solutions of $f(R)$ theories in FLRW spaces are important in the study of early (inflation) and late-time universe (once the matter density is 
sufficiently diluted by the expansion), we hope that the 
universality and simplicity of the proposed method will make it useful in future applications of $f(R)$ theories in cosmology.

{\bf Acknowledgements.} 
This work was in part supported by the project ``Modified gravity theories and the accelerated expansion of the universe'' as a part of the 
Croatian-Serbian bilateral program of cooperation in the field of science and technology. 
S. D. and H. \v{S} acknowledge the support 
by the Ministry of Education, Science and Sports of the Republic of Croatia 
under the contract No. 098-0982930-2864. V. R. acknowledges the support of the Project 171031 of Serbian Ministry of Science.
M. S. is partially supported by the ICTP - SEENET-MTP grant PRJ-09 
(Strings and Cosmology) within the framework of the SEENET-MTP Network. 
M. S. would also like to thank the warm hospitality of the Rudjer Bo\v{s}kovi\'{c} Institute in Zagreb, where part of this work had been done.

\end{document}